# Effect of ballistic re-solution on the nucleation kinetics of precipitates in diluted binary alloys under irradiation. Part 2: Cr-rich α′ precipitates in Fe-Cr alloys

M.S. Veshchunov[*)]

Nuclear Safety Institute (IBRAE), Russian Academy of Sciences,

52, B. Tulskaya, Moscow 115191, Russian Federation

**Abstract**

On the base of a critical analysis of existing models for nucleation of new phase precipitates in metastable binary alloys, a new kinetic model of homogeneous nucleation and growth of α′-phase precipitates in Fe-Cr alloys was developed within the framework of the Reiss kinetic theory of binary nucleation. The model was further modified to account for the influence of ballistic re-solution on the kinetics of nucleation and growth of α′ precipitates, following the general approach proposed in Part 1. The model was used for qualitative interpretation of the results of recent tests, in which the stability of α′ precipitates in Fe-15Cr alloys under irradiation was studied.

[*)] Corresponding author. E-mail: msvesh@gmail.com

## 1. Introduction

Ferrite-martensitic (F-M) steels with high Cr content, which are of technological interest for structural elements of nuclear fusion or fission reactors, are designed to combine corrosion resistance, conferred by chromium, with low swelling and high resistance to irradiation damage, as well as to retain adequate toughness and elevated-temperature strength during service [1]. However, above 12% Cr, F-M steels have been known to undergo hardening and embrittlement during thermal ageing below 748 K due to the nucleation of the $\alpha'$ phase [2], the formation of which is significantly accelerated by neutron irradiation [3, 4].

For these reasons, special attention is paid to the study of binary Fe-Cr model alloys, including specific experimental and theoretical investigations. In particular, recent works address the problem of the numerical simulation of phase separation in these binary alloys, using a combination of density functional theory with statistical approaches involving cluster expansions and Monte Carlo simulations [5, 6]. Also, ab initio calculations and empirical potentials have been used to give new or more realistic physical parameters needed for simulation (e.g., the Fe-rich phase boundary of the $\alpha/\alpha'$ miscibility gap [7]).

These results are a prerequisite for refining analytical models of $\alpha'$ phase nucleation in metastable Fe-Cr alloys, such as the generalised Gibbs nucleation model for two-component solid solutions, considered in the regular solution approximation in [8], which was extended in [9] to include the influence of elastic energy arising from lattice mismatch between the nuclei and the parent phase. In this approach, the basic postulate of the binary nucleation theory formulated by Hobstetter [10] and Reiss [11] was used to calculate the critical nucleus size and composition (and the related nucleation barrier), which for binary systems corresponds to a saddle point on the free energy surface of cluster formation. However, the regular solution models [8, 9] use the same set of microscopic parameters for both phases, which leads to a simplified description of real systems (such as the $\alpha$ and $\alpha'$ phases in Fr-Cr alloy) and, in particular, to a phase diagram that is symmetric with respect to the concentration ½ (which is often not true).

In the present work, this simplification will be overcome by considering a more realistic model of metastable phase decomposition, in which each of the two phases is characterised by its own (independent of the second phase) set of microscopic parameters associated with the actual position of the equilibrium lines (binodals) on the binary phase diagram. For the diluted binary Fe-Cr system with low mutual solubility of the two components in both phases (i.e. Cr in the $\alpha$ phase of the matrix and Fe in the $\alpha'$ phase of the precipitate), each phase can be described in a weak solution approximation. This facilitates the analytical calculation of the composition and size of the critical nucleus, as well as its evolution with the growth of the nucleus.

Furthermore, the nucleation kinetics of binary precipitates is not reduced to simple predictions of classical nucleation theory for unary (single-component) precipitates [12–14] (as assumed in [8]), but should be analysed within the Reiss nucleation kinetic theory [11] modified by Langer [15] (with subsequent reiteration by Stauffer [16]). In the modified theory, the evolution of a cluster through the two-dimensional free energy barrier of nucleation in the vicinity of the critical point does not follow the steepest descent from the saddle point as initially assumed by Reiss (and also considered in [9]), but is determined by the direction of the unstable mode at the saddle point, as shown by Langer [15] (see details in [17]). This approach can be further modified in application to the Fr-Cr alloy (and other similar systems), as proposed in the present work.

In order to avoid difficulties in analytical calculations for the binary systems and to obtain an explicit expression for the nucleation rate, the binary nucleation theory can be simplified using the so called 'quasi-classical' approximation proposed in the author's work [17], which correctly takes into account the direction of cluster growth through the nucleation barrier, while conservatively (under)estimating the pre-exponential kinetic factor of the nucleation rate. To determine this growth direction, it is necessary to analyse the evolution of the composition of the supercritical cluster during its growth, which can be compared with existing experimental observations from the literature.

The model can be further refined to analyse the kinetics of nucleation and growth of $\alpha'$ phase precipitates under irradiation conditions, which have been the subject of detailed study in recent experiments [18]. This will be done based on the general approach proposed in Part 1 for the nucleation of unary phases [19], developing it to consider the behaviour of a binary phase (Cr-rich $\alpha'$ phase) with a small admixture of the second component (Fe).

## 2. Composition and size of the critical nucleus

In accordance with classical nucleation theory [12–14], the Gibbs free energy of a spherical nucleus formation in a binary solid solution takes the form,

$$\Delta G_0(x,y) = x\left(\mu_x^{(p)} - \mu_x^{(m)}\right) + y\left(\mu_y^{(p)} - \mu_y^{(m)}\right) + 4\pi R^2 \gamma, \tag{1}$$

where $x$ and $y$ are the numbers of monomers (atoms) of the two components Fe and Cr, respectively, forming the nucleus; $\mu_x^{(m)}$ and $\mu_y^{(m)}$ are the chemical potentials of these monomers in the matrix; $\mu_x^{(p)}$ and $\mu_y^{(p)}$ are the chemical potentials of these monomers in the spherical cluster (particle) of radius $R = (3/4\pi)^{\frac{1}{3}}(xv_x + yv_y)^{\frac{1}{3}} \approx (3/4\pi)^{\frac{1}{3}}v^{\frac{1}{3}}(x+y)^{\frac{1}{3}}$ with atomic volumes of monomers in the cluster $v_x$ and $v_y$ (which for simplicity are assumed to be equal to each other, as well as to their atomic volumes in the matrix); $\gamma$ is the surface energy of the $\alpha/\alpha'$ interface, which is assumed to be a fixed value (at given temperature), independent of the nucleus composition. A more complex case with the dependence of $\gamma$ on composition and with different atomic volumes of components in the matrix and in the particle (leading to elastic strains in the system) will be discussed below.

In the case of a low concentration of the solute (Cr) in the matrix (Fe), $c_y \ll 1$, the weak solution approximation can be adequately used (cf., e.g. [20]). In this approximation, which takes into account the interaction between solution and solvent components to first (linear) order accuracy, the chemical potentials of the solvent (Fe) and the solute (Cr) are $\mu_x^{(m)} = \mu_x^{(0,m)} - kTc_y$ and $\mu_y^{(m)} = \mu_y^{(0,m)} + kT \ln c_y \equiv kT \ln\left(c_y/c_y^{(0)}\right)$, respectively, where $\mu_x^{(0,m)}$ and $\mu_y^{(0,m)}$ are the standard chemical potentials of the solvent and solute, respectively, and $c_y^{(0)}$ is the thermal concentration of the solute (Cr).

In turn, considering the Cr-rich $\alpha'$ phase as a weak solution of the Fe component in Cr is justified due to the low solubility of Fe in Cr, $n_x \ll 1$, where $n_x = x/(x+y) \approx x/y$ is the Fe concentration in the cluster, and thus $\mu_x^{(p)} = \mu_x^{(0,p)} + kT \ln n_x \equiv kT \ln\left(n_x/n_x^{(0)}\right)$ and $\mu_y^{(p)} = \mu_y^{(0,p)} - kTn_x$, where $\mu_y^{(0,p)}$ and $\mu_x^{(0,p)}$ are the standard chemical potentials of the solvent (Cr) and solute (Fe), respectively, and $n_x^{(0)}$ is the thermal concentration of the solute (Fe). Given that in the weak solution approximation, terms of the second (and higher) order, $c_y^2$ and $n_x^2$, in the free energy functional are ignored, the same accuracy will be applied in further calculations.

From the condition of equilibrium of the two phases (at a flat interface), $\mu_x^{(m)} = \mu_x^{(p)}$ and $\mu_y^{(m)} = \mu_y^{(p)}$, one obtains

$$\mu_x^{(0,m)} = kTc_y^{(eq)} + kT\ln\frac{n_x^{(eq)}}{n_x^{(0)}}, \tag{2}$$

and

$$\mu_y^{(0,p)} = kTn_x^{(eq)} + kT \ln \frac{c_y^{(eq)}}{c_y^{(0)}}, \qquad (3)$$

where $n_x^{(eq)}$ and $c_y^{(eq)}$ are the equilibrium concentrations of the two phases corresponding to binodal lines on the Fe-Cr phase diagram.

Substituting Eq. (2) and (3) in Eq. (1) gives

$$\frac{\Delta G_0(x,y)}{kT} \approx \left\{ x \ln n_x - x \left[ \ln n_x^{(eq)} + 1 - (S-1)c_y^{(eq)} \right] + y \left( n_x^{(eq)} - \ln S \right) \right\} + \frac{(36\pi)^{\frac{1}{3}} \gamma v^{\frac{2}{3}}}{kT} (x+y)^{\frac{2}{3}}, \qquad (4)$$

where $S = c_y / c_y^{(eq)}$ is the oversaturation ratio of Cr atoms in the Fe matrix.

According to Reiss' binary nucleation theory, the critical saddle point $(x^*, y^*)$ is determined by minimization of the formation free energy, Eq. (4), with respect to the two variables, $\partial \Delta G_0(x,y)/\partial x|_{x^*,y^*} = \partial \Delta G_0(x,y)/\partial y|_{x^*,y^*} = 0$, which leads to the following equations for the critical nucleus,

$$\ln n_x - \ln n_x^{(eq)} + (S-1)c_y^{(eq)} + \frac{2}{3}(36\pi)^{\frac{1}{3}} \frac{\gamma v^{\frac{2}{3}}}{kT} (x+y)^{-\frac{1}{3}} = 0, \qquad (5)$$

and

$$-n_x + n_x^{(eq)} - \ln S + \frac{2}{3}(36\pi)^{\frac{1}{3}} \frac{\gamma v^{\frac{2}{3}}}{kT} (x+y)^{-\frac{1}{3}} = 0, \qquad (6)$$

superposition of which gives,

$$\ln \frac{n_x}{n_x^{(eq)}} = -n_x + n_x^{(eq)} - (S-1)c_y^{(eq)} - \ln S, \qquad (7)$$

or,

$$z \equiv n_x \exp n_x = \frac{n_x^{(eq)}}{S} \exp \left[ n_x^{(eq)} - (S-1)c_y^{(eq)} \right] \ll 1, \qquad (8)$$

which solution determines the critical cluster composition $n_x^*$.

From Eq. (8) it is seen that $n_x^* \to n_x^{(eq)}$ as $S \to 1$, whereas taking into account that

$$\frac{dz}{dS} = \frac{d}{dS} \left\{ \frac{n_x^{(eq)}}{S} \exp \left[ n_x^{(eq)} - (S-1)c_y^{(eq)} \right] \right\} = -\frac{n_x^{(eq)}}{S} \left( \frac{1}{S} + c_y^{(eq)} \right) \exp \left[ n_x^{(eq)} - (S-1)c_y^{(eq)} \right] < 0, \qquad (9)$$

one obtains that $z(S \geq 1) \leq z(S = 1) = n_x^{(eq)} \exp n_x^{(eq)}$, or $n_x^* \exp n_x^* \leq n_x^{(eq)} \exp n_x^{(eq)}$, or $n_x^* \leq n_x^{(eq)}$.

The solution of Eq. (8) is

$$n_x^* = W \left\{ \frac{n_x^{(eq)}}{S} \exp \left[ n_x^{(eq)} - (S-1)c_y^{(eq)} \right] \right\} = W(z), \qquad (10)$$

where $W(z)$ is the Lambert function, which decomposes at small $z$ as $W(z) \approx z - z^2 + \frac{3}{2}z^3 + \cdots$, and hence with accepted linear accuracy,

$$n_x^* \approx \frac{n_x^{(eq)}}{S} \exp \left[ n_x^{(eq)} - (S-1)c_y^{(eq)} \right] \approx \frac{n_x^{(eq)}}{S}. \qquad (11)$$

Substituting Eq. (11) into Eq. (6), in the first approximation for $n_x^{(eq)} \ll 1$, leads to

$$y^* \approx \frac{32\pi\gamma^3 v^2}{3(kT)^3} \frac{1}{\left(\ln S - n_x^{(eq)} + n_x^*\right)^3} \frac{1}{(1+n_x^*)} \approx \frac{32\pi\gamma^3 v^2}{3(kT)^3} \frac{1}{\left(\ln S - n_x^{(eq)} + n_x^*\right)^3}, \tag{12}$$

and

$$x^* \approx y^* n_x^* \approx \frac{32\pi\gamma^3 v^2}{3(kT)^3} \frac{n_x^*}{\left(\ln S - n_x^{(eq)} + n_x^*\right)^3}. \tag{13}$$

Substituting of Eqs (11)–(13) into Eq. (4) gives for the nucleation barrier,

$$\frac{\Delta G_0(x^*, y^*)}{kT} \approx \frac{16\pi}{3} \frac{\gamma^3 v^2}{(kT)^3 \left(\ln S - n_x^{(eq)} + n_x^*\right)^2} \approx \frac{16\pi}{3} \frac{\gamma^3 v^2}{(kT)^3 \left[\ln S - n_x^{(eq)} + \frac{n_x^{(eq)}}{S}\right]^2}, \tag{14}$$

which can be simplified at small supersaturations, $S - 1 \approx \ln S \ll 1$, as

$$\frac{\Delta G_0(x^*, y^*)}{kT} \approx \frac{16\pi}{3} \frac{\gamma^3 v^2}{(kT)^3 \left[\left(1 - \frac{n_x^{(eq)}}{S}\right) \ln S\right]^2}. \tag{15}$$

It is seen that the critical supersaturation $S^*$, above which the nucleation of the new phase becomes possible and which is determined by the pole of Eq. (15), is not shifted, $S^* = 1$, whereas the critical nucleus composition is below the binodal concentration, $n_x^* \le n_x^{(eq)}$ (as shown above), i.e. the composition of the critical nucleus belongs to the stable zone of the $\alpha'$ phase on the Fe-Cr phase diagram.

The latter conclusion can be verified by taking into account the composition dependence of the interface energy $\gamma$ using the quasi-chemical treatment of Becker [21] and Turnbull [22], in which the interface energy corresponds to the difference between the energies of bonds broken in the separation process and of bonds made in forming the interphase boundary, leading to

$$\gamma = \gamma_0 (c_x - n_x)^2 = \gamma_0 (1 - c_y - n_x)^2, \tag{16}$$

as considered in [23].

In this approach, Eqs (5) and (6) are modified by the additional terms, $(36\pi)^{\frac{1}{3}} v^{\frac{2}{3}} (x + y)^{\frac{2}{3}} \frac{d\gamma}{kTdx} = \frac{3v}{kTR} \frac{d\gamma}{dx} (x + y) = \frac{3v}{kTR} \frac{d\gamma}{dn_x} (1 - n_x)$ and $(36\pi)^{\frac{1}{3}} v^{\frac{2}{3}} (x + y)^{\frac{2}{3}} \frac{d\gamma}{kTdy} = \frac{d\gamma}{dy} \frac{3v}{kTR} (x + y) = -\frac{3v}{kTR} \frac{d\gamma}{dn_x} n_x$, respectively, where, as seen from Eq. (16), $\frac{d\gamma}{dn_x} = -\gamma_0 n_x (1 - c_y - n_x) < 0$.

Accordingly, Eq. (7) takes the form,

$$\ln \frac{n_x}{n_x^{(eq)}} = -n_x + n_x^{(eq)} - (S-1)c_y^{(eq)} - \ln S + 3n_x(1 - c_y - n_x)\frac{\gamma_0 v}{RkT}, \tag{17}$$

which can be simplified to first order accuracy for $n_x, c_y \ll 1$ as

$$\ln \frac{n_x}{n_x^{(eq)}} \approx -n_x \left(1 - 3\frac{\gamma_0 v}{RkT}\right) + n_x^{(eq)} - (S-1)c_y^{(eq)} - \ln S, \tag{18}$$

or,

$$n_x \left(1 - 3\frac{\gamma_0 v}{RkT}\right) + \ln\left[n_x \left(1 - 3\frac{\sigma_0 v}{RkT}\right)\right] \approx n_x^{(eq)} + \ln\left[n_x^{(eq)} \left(1 - 3\frac{\gamma_0 v}{RkT}\right)\right] - (S-1)c_y^{(eq)} - \ln S, \tag{19}$$

whose solution is

$$n_x^* \left(1 - 3\frac{\gamma_0 v}{RkT}\right) \approx W\left\{\frac{n_x^{(eq)}}{S} \left(1 - 3\frac{\gamma_0 v}{R^*kT}\right) \exp\left[n_x^{(eq)} - (S-1)c_y^{(eq)}\right]\right\}, \tag{20}$$

where $0 < \left(1 - 3\frac{\gamma_0 v}{R^* kT}\right) < 1$ for the typical values $\gamma_0 \approx 0.2$ J·m$^{-2}$ (evaluated in [24]), $T \approx 600$–700 K, and $R^* \geq 1$ nm. Therefore, Eq. (20) can be reduced to

$$n_x^* \approx \frac{n_x^{(eq)}}{S}\left(1 - 3\frac{\gamma_0 v}{R^* kT}\right) \exp\left[n_x^{(eq)} - (S-1)c_y^{(eq)}\right], \tag{21}$$

which is smaller than the expression in Eq. (11) (due to the additional factor $\left(1 - 3\frac{\gamma_0 v}{R^* kT}\right) < 1$), i.e. the composition of the critical nucleus penetrates even deeper into the stable zone of the $\alpha'$ phase on the phase diagram. For the above parameters, this factor is quite sensitive to the value of $\gamma_0$, which contradicts the results of the model [9], which also used Eq. (16) but predicted the independence of the critical nucleus composition from $\gamma_0$ (as specified in [9]).

On the other hand, Eq. (21) contradicts the results of the earlier model (for the nucleation Cu-rich $\varepsilon$ phase in a binary Fe-Cu alloy) [23], which, using Eq. (16), predicted a much less Cu in the critical nucleus compared to the equilibrium $\varepsilon$ phase (i.e. the composition of the critical nucleus was found to be deep within miscibility gap of the phase diagram).

For a coherent spherical particle with the misfit strain $\delta = \Delta a/a = (a' - a)/a \ll 1$, where $a$ and $a'$ are the lattice constants in the matrix and particle, respectively, the elastic strain energy, which contributes to the cluster formation energy, Eq. (4), can be estimated in the elastically homogeneous approximation as $\Delta G_{el} = \frac{E}{2(1-\nu)}\delta^2 V_p$ (cf. [25]), where $E \approx 200$ GPa is the Young modulus, $\nu \approx 1/3$ is the Poisson ratio, and $\delta \approx 10^{-2}$ at the $\alpha/\alpha'$ interface. Taking into account similar crystal structures of the two phases, Vegard's law can be used for evaluation of the misfit strain as $\Delta a = a' - a = a\delta = (a_{Fe} - a_{Cr})(1 - c_y - n_x)$ [9].

In this approach, the l.h.s. of Eqs (5) and (6) are modified by the additional terms, $-\frac{2\Gamma v}{kT}(1 - c_y - n_x)(1 + n_x) + \frac{\Gamma v}{kT}(1 - c_y - n_x)^2$ and $\frac{2\Gamma v}{kT}(1 - c_y - n_x)n_x + \frac{\Gamma v}{kT}(1 - c_y - n_x)^2$, respectively, where $\Gamma \equiv \frac{Ev}{2(1-\nu)}\delta^2$, which replaces Eq. (7) with

$$\ln n_x - \ln n_x^{(eq)} + (S-1)c_y^{(eq)} - \frac{2\Gamma v}{kT}(1 - c_y - n_x)(1 + n_x) = -n_x + n_x^{(eq)} - \ln S + \frac{2\Gamma v}{kT}(1 - c_y - n_x)n_x, \tag{22}$$

or, to first order accuracy for $n_x, c_y \ll 1$,

$$n_x\left[1 - \frac{2\Gamma v}{kT}(1 - 2c_y)\right] + \ln n_x \approx n_x^{(eq)} + \ln n_x^{(eq)} - \ln S - (S-1)c_y^{(eq)} + \frac{2\Gamma v}{kT}(1 - c_y), \tag{23}$$

whose solution is

$$n_x\left[1 - \frac{2\Gamma v}{kT}(1 - 2c_y)\right] = W\left\{\frac{n_x^{(eq)}}{S}\left(1 - \frac{2\Gamma v}{kT}(1 - 2c_y)\right)\exp\left[n_x^{(eq)} - (S-1)c_y^{(eq)} + \frac{2\Gamma v}{kT}(1 - c_y)\right]\right\}, \tag{24}$$

which reduces to

$$n_x \approx \frac{n_x^{(eq)}}{S}\exp\left[n_x^{(eq)} - (S-1)c_y^{(eq)}\right]\exp\left[\frac{2\Gamma v}{kT}(1 - c_y)\right]. \tag{25}$$

This solution differs from the expression in Eq. (11) by the factor $\exp[2\Gamma v(1 - c_y)/kT] \approx 1$, (given that $2\Gamma v/kT \approx 6 \cdot 10^{-2} \ll 1$), which has practically no effect on the composition of the critical cluster (contrary to the predictions of the regular solution model [9]).

## 3. Growth of precipitates

In the binary Fe-Cr system, the growth of the nucleated $\alpha'$-phase clusters can be described by a system of two rate equations for the Cr and Fe components.

The rate of the increase in the number of Cr atoms in the precipitate, $\dot{y}$, can be calculated using the balance equation,

$$\dot{y} = 4\pi R^2 \left[\beta_y(x,y) - \alpha_y^{(th)}(x,y)\right] = 4\pi D S c_y^{(eq)} v^{-1} R \left[1 - \exp\left(\frac{1}{kT}\frac{\partial \Delta G_0(x,y)}{\partial y}\right)\right], \tag{26}$$

where $\beta_y(x,y) = D c_y R^{-1} v^{-1}$ is the arrival rate of Cr atoms by diffusion from the matrix with the Cr concentration $c_y$; $\alpha_y^{(th)}(x,y) = \beta_y(x,y) \exp\left(\frac{1}{kT}\frac{\partial \Delta G_0(x,y)}{\partial y}\right)$ is the thermal loss rate, determined from the Gibbs-Kelvin condition of equilibrium between Cr atoms in the particle and in the matrix.

The rate of the increase in the number of Fe atoms in the precipitate, $\dot{x}$, can be calculated as

$$\dot{x} = 4\pi R^2 \left[\beta_x(x,y) - \alpha_x^{(th)}(x,y)\right], \tag{27}$$

where $\beta_x(x,y) \approx D(1 - c_y) a^{-1} v^{-1}$ is the arrival rate of Fe atoms from the matrix with the Fe concentration $c_x = 1 - c_y$, which is determined by atomic jumps over a distance $a$ at the interface, taking into account a linear correction (for small $c_y \ll 1$) to the arrival rate from the pure Fe matrix; $\alpha_x^{(th)}(x,y) = \beta_x(x,y) \exp\left(\frac{1}{kT}\frac{\partial \Delta G_0(x,y)}{\partial x}\right)$ is the thermal loss rate, determined from the Gibbs-Kelvin condition of equilibrium between Fe atoms in the particle and in the matrix.

Given that $\frac{\beta_x}{\beta_y} = \frac{(1-c_y)}{c_y}\frac{R}{a} \sim 10^2 \gg 1$, the condition for the finite ratio of these rates, $\frac{\dot{x}}{\dot{y}} \approx \frac{\beta_x}{\beta_y}\left[1 - \exp\left(\frac{1}{kT}\frac{\partial \Delta G_0(x,y)}{\partial x}\right)\right] / \left[1 - \exp\left(\frac{1}{kT}\frac{\partial \Delta G_0(x,y)}{\partial y}\right)\right] \sim \frac{x}{y} \sim n_x^{(eq)} \ll 1$, which ensures the mixed composition of the growing particle, corresponds to the adiabatic condition, $\partial \Delta G_0(x,y)/\partial x \approx 0$, which can replace Eq. (27) with sufficient accuracy.

Therefore, in this (adiabatic) approximation, the composition of the growing precipitate obeys the equation (cf. Eq. (5)),

$$n_x \approx n_x^{(eq)} \exp\left[-(S-1)c_y^{(eq)} - \frac{2\gamma v}{kTR}\right], \tag{28}$$

which determines $x(y) = n_x y$, and when substituted into Eq. (4), gives an expression for the free energy of the growing precipitate,

$$\Delta G_0(y) \equiv \Delta G_0(x(y),y) \approx kT \left\{ y\left(n_x^{(eq)} - \ln S\right) - y n_x^{(eq)}\left(1 + \frac{2\gamma v}{kTR}\right)\exp\left[-(S-1)c_y^{(eq)} - \frac{2\gamma v}{kTR}\right]\right\} + \gamma(36\pi)^{\frac{1}{3}} v^{\frac{2}{3}} y^{\frac{2}{3}} \left[1 + n_x^{(eq)}\exp\left[-(S-1)c_y^{(eq)} - \frac{2\gamma v}{kTR}\right]\right]^{\frac{2}{3}}, \tag{29}$$

leading to

$$\frac{1}{kT}\frac{d\Delta G_0(y)}{dy} \approx n_x^{(eq)} - \ln S + \frac{2\gamma v}{kTR} + n_x^{(eq)}\frac{2\gamma v}{kTR}\exp\left[-(S-1)c_y^{(eq)} - \frac{2\gamma v}{kTR}\right], \tag{30}$$

where $R = (3/4\pi)^{\frac{1}{3}} v^{\frac{1}{3}} y^{\frac{1}{3}} (1 + n_x(y))^{\frac{1}{3}} \approx (3/4\pi)^{\frac{1}{3}} v^{\frac{1}{3}} y^{\frac{1}{3}}$, taking into account that $n_x(y) \ll 1$ (as will be shown below).

Consequently, the precipitate growth rate, $\dot{y}$, can be calculated using the balance equation,

$$\dot{y} = 4\pi R^2 [\beta(y) - \alpha^{(th)}(y)] = 4\pi D S c_y^{(eq)} v^{-1} R \left[1 - \exp\left(\frac{1}{kT}\frac{d\Delta G_0(y)}{dy}\right)\right], \tag{31}$$

which in the first approximation for $n_x^{(eq)}, c_y^{(eq)} \ll 1$, for relatively large particles, $R \gg 2\gamma v/kT \approx 0.5$ nm, using Eq. (30) gives,

$$\dot{y} \approx \frac{d}{dt}\left(\frac{4\pi}{3v}R^3\right) \approx 4\pi D v^{-1} R \left[c_y - c_y^{(eq)} \exp\left(n_x^{(eq)}\right) \exp\left(\frac{2\gamma v}{kTR}\right)\right]. \tag{32}$$

With an increase in particle radius, the particle composition $n_x$, Eq. (28), also increases, and for large particles, $R \gg \frac{2\gamma v}{kT(S-1)c_y^{(eq)}}$, becomes close to the equilibrium concentration of the $\alpha'$ phase,

$$n_x(y) \to n_x^{(eq)} \exp\left[-(S-1)c_y^{(eq)}\right] \approx n_x^{(eq)}. \tag{33}$$

This model prediction is in reasonable agreement with Mössbauer spectroscopy measurements in long term ageing tests [26], where the Cr concentration in $\alpha'$-phase precipitates at 748 K was 86 at.%, which is quite close to the equilibrium concentration of $\approx 88$ at.% (according to the Calphad database [27]). Similar results were obtained by atom probe tomography (APT) in a series of model Fe‑Cr alloys containing more than 9 at.% Cr and irradiated by neutrons to 1.82 dpa at a nominal temperature of 563 K [28], suggesting irradiation-accelerated precipitation, where the averaged Cr concentration in $\alpha'$-phase precipitates was consistent with the phase diagram. The discrepancy with earlier investigations of thermally aged Fe-20%Cr alloys at $T = 773$ K by APT [29], where the composition of $\alpha'$-phase precipitates evolved temporally from $(60.6 \pm 0.9)$ at.% after 50 h up to $(83 \pm 1)$ at.% after 1067 h, was analysed in [28], where it was associated with a very strong magnification effect (where Cr has a lower evaporation field than Fe) for small nanoscale precipitates formed during short ageing periods. In addition, it might be assumed that at the relatively high test temperature [29], corresponding to the eutectoid decomposition of the $\sigma$ phase [30], precipitates with a Cr concentration of $\approx 50$ at.% could have formed, which affected measurement results at short ageing times.

Another possible reason for this discrepancy may be related to the high degree of alloy supersaturation in these tests [29], when the Gibbs free energy of nucleus formation, $\Delta G_0(x, y)$, cannot be described using the weak solution approximation (i.e. Eq. (4)). In particular, this approximation cannot account for nonlinear effects, which become important for describing phase transitions in metastable systems with compositions approaching the spinodal line.

Besides, for such systems with a high degree of metastability, classical nucleation theory may prove insufficient, and consideration within the framework of the Van der Waals theory may be required (as discussed in [8]). This theory takes into account continuous changes in thermodynamic parameters at the interface and their dependence on concentration gradients [31, 32]. A similar approach was used in phase field calculations [33], where the evolution of the composition of Cr-rich precipitates was found to be comparable with the results of experiments [29]. In particular, it was shown that effects of non-classical nucleation theory begin to manifest themselves for alloys with Cr concentration above $\approx 0.17$.

### 4. Nucleation rate of precipitates

Classical nucleation theory [12–14] was developed in relation to nucleation of single-component (unary) phases, and for this reason, is not directly applicable to nucleation of the $\alpha'$ phase, which occurs by agglomeration of two components (Fe and Cr) in precipitates.

The nucleation theory was generalized to two-component (binary) systems by Reiss [11]. In his theory, the parent phase is thought of as a mixture of molecules (monomers) of two components X and Y with number densities $N_x$ and $N_y$, respectively, together with clusters of all sizes and compositions. A particular molecular cluster is characterized by the numbers of single molecules (or monomers) $x$ and $y$ of species X and Y, respectively, that it contains. Reiss showed that the critical point of unstable equilibrium (associated with the phase transition) corresponds in this case to a saddle point $(x^*, y^*)$ on the free energy surface $\Delta G_0(x, y)$. He characterized the transition by a two-dimensional steady state flow $J(x, y)$ of clusters in the phase space of cluster sizes $(x, y)$, which is pronounced in one direction (the axis of the pass $x'$) that, in comparison with it, any lateral flow (in the perpendicular direction $y'$) may be neglected, i.e. $J_{y'} \approx 0$. Due to the steady state condition, $\text{div} J = \frac{\partial J_{x'}}{\partial x'} + \frac{\partial J_{y'}}{\partial y'} \approx \frac{\partial J_{x'}}{\partial x'} = 0$, this leads to $J_{x'} \approx J(y')$, which was calculated by Reiss as

$$J(y' - y^*) = f_0(x^*, y^*) \frac{\beta_x^* \beta_y^* (1 + \tan^2 \theta)}{\beta_y^* + \beta_x^* \tan^2 \theta} \left( \frac{|D'_{11}|}{\pi kT} \right)^{1/2} \exp\left[ -\frac{|\det \mathbf{D}|(y' - y^*)^2}{kT|D'_{11}|} \right], \tag{34}$$

where $\rho_0(x, y)$ is the equilibrium size distribution function,

$$\rho_0(x, y) = F \exp[-\Delta G_0(x, y)/kT], \tag{35}$$

$F = N_y = c_y v^{-1}$ is the so called number density of potential nucleation sites, determined for binary alloys according to the Frenkel model [34] (additionally justified in the Appendix C to [35]); $\theta$ is the angle between the original axis $x$ and the axis of the pass $x'$; $\beta_i^* = \beta_i(x^*, y^*)$, $i = x, y$, are the arrival rates of monomers X and Y to the critical cluster $(x^*, y^*)$ of radius $R^*$; $D_{ij} = (1/2)\, \partial^2 \Delta G_0(x^*, y^*)/\partial x_i \partial x_j$ are elements of the matrix $\mathbf{D} = (D_{ij})$, which determinant is negative (in accordance with the properties of the saddle point), $\det \mathbf{D} = D_{11} D_{22} - D_{12}^2 < 0$;

$$D'_{11} = \frac{1}{2} \frac{\partial^2 \Delta G_0(x', y')}{\partial x'^2} \bigg|_{x^*, y^*} = D_{11} \cos^2 \theta + D_{22} \sin^2 \theta + 2 D_{12} \sin \theta \cos \theta, \tag{36}$$

is the second derivative of $\Delta G_0$ at the critical point in the direction $x'$ of the orthogonal coordinate system $(x', y')$ obtained by rotating the original coordinate system $(x, y)$ through the angle $\theta$; this derivative should be negative, $D'_{11} < 0$, to provide a maximum of the free energy at the critical point in the direction of the $x'$-axis.

Therefore, the nucleation rate, defined as the total flux of clusters through the critical zone,

$$\dot{N} = \int_{-\infty}^{\infty} J(y' - y^*) dy', \tag{37}$$

was calculated by Reiss by substituting Eq. (34) into Eq. (37) as

$$\dot{N} \approx -\rho_0(x^*, y^*) \frac{\beta_x^* \beta_y^* (1 + \tan^2 \theta)}{\beta_y^* + \beta_x^* \tan^2 \theta} D'_{11} \left( \frac{1}{D_{12}^2 - D_{11} D_{22}} \right)^{1/2}. \tag{38}$$

In the Reiss theory, the axis of the pass $x'$ runs in the direction of the steepest descent of the free energy surface $\Delta G_0(x, y)$, which for this reason was determined in the thermodynamic approach (i.e. solely from the properties of the free energy). This assumption was modified by Langer [15] (with subsequent reiteration by Stauffer [16]), who corrected the orientation of the flux vector in the direction parallel to the direction of the unstable mode at the saddle point (the new axis of the pass $x'$). The modified value of $\theta$ was explicitly calculated in [16] and later refined in [36] as

$$\tan \theta = s + (r + s^2)^{1/2}, \quad \text{if } D_{21} < 0, \tag{39}$$

and

$$\tan \theta = s - (r + s^2)^{1/2}, \quad \text{if } D_{21} > 0, \tag{40}$$

where $r = \beta_y^* / \beta_x^*$, $s = (d_x - r d_y)/2$, $d_x = -D_{11}/D_{12}$ and $d_y = -D_{22}/D_{12}$.

The parameters of these equations are evaluated in the Appendix, resulting in an analytical expression for the nucleation rate of the $\alpha'$ phase, which however is rather awkward for practical applications. Therefore, it can be simplified using the so called 'quasi-classical' approximation proposed in the author's work [17].

*4.1. Quasi-classical approximation*

The total flow contribution of off-critical nuclei with higher formation barriers (compared to the critical nucleus $(x^*, y^*)$), which move along trajectories adjacent to the 'classical' trajectory $x_c(y)$ (describing the critical cluster evolution in the supercritical zone) can be considered as a correction, which is neglected in the quasi-classical approximation (taking into account only the extreme classical trajectory). Correspondingly, only the contribution of the flux $J(y' = y^*)$ to the integral in Eq. (37) is taken into account. This consideration can be substantiated in the case when the saddle point passage is narrow and has very steep sides, and only those clusters with the composition of the saddle point nucleate in any noticeable quantity. This formally corresponds to the condition $|\det \mathbf{D}|/|D'_{11}| \gg \pi kT$, which is fulfilled in the case of $\alpha'$-phase precipitates if $n_x^{(eq)} \ll 0.1$, whereas for $n_x^{(eq)} \approx 0.1$ it can be considered as an approximation that conservatively estimates the nucleation rate.

In this approximation, the nucleation problem becomes quasi-one-dimensional, by formally substituting $x_c(y)$ into the expression for $\Delta G_0(x, y)$ and further analysing a one-dimensional system with $\Delta G_0(y) = \Delta G_0(x_c(y), y)$ in the framework of classical nucleation theory. In application to the Fe-Cr clusters, the classical trajectory can be calculated using Eq. (28) as $x_c(y) = y n_x(y)$, whereas $\Delta G_0(y)$ is described by Eq. (29).

This makes it possible to carry out a simplified analysis, in which the nucleation rate of the binary Fe-Cr precipitates is calculated as

$$\dot{N} \approx \omega^* \rho_0(y^*) Z, \tag{41}$$

where $y^*$ is the critical nucleus size determined by the extremum of $\Delta G_0(y)$; $\omega^*$ is the arrival rate of monomers (Cr atoms) to the critical nucleus;

$$\rho_0(y^*) = N_y \exp[-\Delta G_0(y^*)/kT] = S c_y^{(eq)} v^{-1} \exp[-\Delta G_0(x^*, y^*)/kT], \tag{42}$$

is the value of the equilibrium size distribution function for the critical nucleus, and

$$Z = \left[-\frac{1}{2\pi kT} \frac{d^2 \Delta G_0(y^*)}{dy^2}\right]^{\frac{1}{2}}, \tag{43}$$

is the Zeldovich factor.

For supercritical clusters with radius close to the critical radius, $R \approx R^*$ (and thus $\partial \Delta G_0(y)/\partial y \to 0$), Eq. (31) can be simplified as

$$\frac{dy}{dt} \approx -4\pi D R(y) S c_y^{(eq)} v^{-1} \frac{1}{kT} \frac{\partial \Delta G_0(y)}{\partial y}, \tag{44}$$

which, using the general relationship of Zeldovich's theory [14] (see also [37]), allows calculating the arrival rate to the critical nucleus as

$$\omega^* = -kT \left(\frac{dy}{dt}\right) \left(\frac{\partial \Delta G_0(y)}{\partial y}\right)^{-1} \bigg|_{y=y^*} = 4\pi D R^* S c_y^{(eq)} v^{-1} = 4\pi D S c_y^{(eq)} \frac{2\gamma}{kT} \frac{1}{\ln S - n_x^{(eq)} + n_x^*}. \tag{45}$$

The Zeldovich factor is calculated using Eq. (30) as

$$\frac{1}{kT} \frac{d^2 \Delta G_0(y)}{dy^2} \approx -\frac{2\gamma v}{3kTRy} \left\{1 - n_x^{(eq)} \left(1 - \frac{2\gamma v}{kTR}\right) n_x^{(eq)} \exp\left[-(S-1) c_y^{(eq)} - \frac{2\gamma v}{kTR}\right]\right\}, \tag{46}$$

which, after substituting into Eq. (43) and using Eq. (12), gives

$$Z \approx \frac{1}{8\pi} \frac{(kT)^{3/2}}{\gamma^{3/2} v} \left[\ln S - n_x^{(eq)}(1 - S^{-1})\right]^2 \left\{1 + \frac{1}{3}\left[\ln S - n_x^{(eq)}(1 - S^{-1})\right] n_x^{(eq)} S^{-1} \exp\left[-(S - 1)c_y^{(eq)} + n_x^{(eq)}(1 - S^{-1})\right]\right\}^{\frac{1}{2}}, \quad (47)$$

or, in the first approximation for small parameters $n_x^{(eq)}, c_y^{(eq)} \ll 1$,

$$Z \approx \frac{1}{8\pi} \frac{(kT)^{3/2}}{\gamma^{3/2} v} \left[\ln S - n_x^{(eq)}(1 - S^{-1})\right]^2. \quad (48)$$

Substituting Eqs (42), (45), (48) and (14) into Eq. (41), the nucleation rate can be calculated as

$$\dot{N} \approx D\left(Sc_y^{(eq)}\right)^2 \frac{(kT)^{\frac{1}{2}}}{\gamma^{\frac{1}{2}} v^2} \left[\ln S - n_x^{(eq)}(1 - S^{-1})\right] \exp\left\{-\frac{16\pi}{3} \frac{\gamma^3 v^2}{(kT)^3 \left[\ln S - n_x^{(eq)}(1 - S^{-1})\right]^2}\right\}, \quad (49)$$

which is a conservative estimate that can be improved by using a more accurate Eq. (38) (with parameters estimated in the Appendix).

## 5. Precipitation under irradiation conditions

A modification of classical nucleation theory for unary precipitates in metastable solid solutions under irradiation, taking into account the influence of ballistic re-solution on the nucleation kinetics of precipitates (consisting of $y$ monomers), was carried out in Part 1 [19]. In particular, it was shown that under steady state irradiation, the nucleation rate is determined by a 'quasi-equilibrium' size distribution function, $\rho_0(y) = C \exp[-\Delta\bar{G}_0(y)/kT]$, where $\Delta\bar{G}_0(y)$ is a non-thermodynamic function that is a 'quasi-equilibrium' analogue of the free energy of cluster formation in the absence of irradiation.

As a result, it was shown that $\Delta\bar{G}_0(y)$ can have two extrema. The first is an unstable extremum that defines the critical cluster radius, and the second is a stable extremum that defines the maximum particle radius at which supercritical particle growth ceases. Accordingly, these extrema corresponds to the condition of zero growth rate of a particle. With a decrease in the supersaturation $S$, the two extrema converge until they merge into a single radius $\hat{R}$, reached at the threshold supersaturation $\hat{S}$, below which nucleation of precipitates is unfeasible. In particular, this threshold effect may explain the suppression of new phase formation, if the supersaturation in the solid solution is below the threshold value, $1 \leq S < \hat{S}$.

This theory can be applied, in the quasi-classical approximation (described above in Section 4.1), to the nucleation of the $\alpha'$ phase in Fe-Cr alloys by introducing a ballistic re-solution term into the rate equation for Cr atoms, Eq. (26),

$$\dot{y} \approx 4\pi R^2 \left[\beta_y(x,y) - \alpha_y^{(th)}(x,y) - \alpha_y^{(irr)}(x,y)\right] = 4\pi D Sc_y^{(eq)} v^{-1} R \left[1 - \exp\left(\frac{1}{kT}\frac{\partial \Delta G_0(x,y)}{\partial y}\right) - \frac{R\xi K_0(1-n_x)}{DSc_y^{(eq)}}\right], \quad (50)$$

where $\alpha_y^{(irr)}(x,y) = (1 - n_x)\xi K_0 v^{-1}$ is the rate of ballistic re-solution of Cr atoms from precipitates, $K_0$ is the damage rate in the irradiated solid, and $\xi$ is the irradiation re-solution constant evaluated in [38] as $\approx 15$ nm. Similarly, the rate equation for Fe atoms, Eq. (27), is modified to the form,

$$\dot{x} = 4\pi R^2 \left[ \beta_x(x,y) - \alpha_x^{(th)}(x,y) - \alpha_x^{(irr)}(x,y) \right] =$$
$$4\pi R^2 D(1-c_y)a^{-1}v^{-1}\left[1 - \exp\left(\frac{1}{kT}\frac{\partial \Delta G_0(x,y)}{\partial x}\right) - \frac{n_x \xi K_0 a}{D(1-c_y)}\right], \tag{51}$$

where $\alpha_x^{(irr)}(x,y) = n_x \xi K_0 v^{-1}$ is the rate of ballistic re-solution of Fe atoms from precipitates.

The precipitate composition can be determined from the adiabatic condition for Fe atoms (as explained above in Section 3), which in this case takes the form,

$$n_x \left[1 + \frac{n_x \xi K_0 a}{D(1-c_y)}\right] \approx n_x^{(eq)} \exp\left[-(S-1)c_y^{(eq)} - \frac{2\gamma v}{kTR}\right], \tag{52}$$

and can replace Eq. (51) with sufficient accuracy. Taking into account that $\Psi \equiv \xi K_0 a/D \sim 10^{-3} - 10^{-2} \ll 1$ (using values of $D/K_0 \approx 10^{-15} - 10^{-14}$ m²/s, estimated below in Section 6), Eq. (52) can be reduced to Eq. (28) with sufficient accuracy.

Therefore, after substituting Eq. (28) into Eq. (50), the growth rate equation takes the form,

$$\frac{d}{dt}\left(\frac{4\pi}{3v}R^3\right) = \dot{y} + \dot{x} = s(y)\left[\tilde{\beta}_y(y) - \tilde{\alpha}_y^{(th)}(y) + \tilde{\alpha}_y^{(irr)}(y)\right], \tag{53}$$

where $\tilde{\beta}_y \equiv \beta_y(x(y),y)$, $\tilde{\alpha}_y^{(th)}(y) \equiv \alpha_y^{(th)}(x(y),y)$, $\tilde{\alpha}_y^{(irr)}(y) \equiv \alpha_y^{(irr)}(x(y),y)$, $s(y) = 4\pi R^2(y)$, which, in the first approximation for $n_x^{(eq)}, c_y^{(eq)}, \Psi/c_y^{(eq)} \ll 1$ and $R \gg 2\gamma v/kT \approx 0.5$ nm (and thus $n_x \sim n_x^{(eq)} \ll 1$, as seen from Eq. (52)), modifies Eq. (32) to the form,

$$\dot{R} \approx DR^{-1}\left[S - \exp\left(\frac{2\gamma v}{kTR}\right)\exp\left(n_x^{(eq)}\right) - \frac{\xi K_0}{Dc_y^{(eq)}}R\right] \approx DR^{-1}\left[S - 1 + n_x^{(eq)} + \frac{2\gamma v}{kTR} - \frac{\xi K_0}{Dc_y^{(eq)}}R\right]. \tag{54}$$

Accordingly, the condition of zero growth rate, $\dot{R} = 0$, takes the form,

$$S - 1 - n_x^{(eq)} - \frac{2\gamma v}{kTR} - \frac{\xi K_0}{Dc_y^{(eq)}}R \approx 0, \tag{55}$$

which has two roots,

$$R_{1,2}^* = \frac{Dc_y^{(eq)}(S-1-n_x^{(eq)})}{2\xi K_0}\left\{1 \mp \left[1 - \frac{8\gamma v \xi K_0}{kTDc_y^{(eq)}(S-1-n_x^{(eq)})^2}\right]^{1/2}\right\}, \tag{56}$$

existing if the discriminant is positive. Accordingly, the condition of zero determinant is used to determine the threshold supersaturation below which no solutions exist,

$$\hat{S} \approx 1 + n_x^{(eq)} + \left(\frac{8\gamma v \xi K_0}{kTDc_y^{(eq)}}\right)^{\frac{1}{2}}, \tag{57}$$

which, after substituting into Eq. (56), defines the threshold radius,

$$\hat{R} = \frac{Dc_y^{(eq)}(\hat{S}-1-n_x^{(eq)})}{2\xi K_0} \approx \left(\frac{2\gamma v Dc_y^{(eq)}}{kT\xi K_0}\right)^{\frac{1}{2}}. \tag{58}$$

The nucleation rate under irradiation conditions can be calculated using the Fuchs algorithm for determining the quasi-equilibrium size distribution function (see Part 1), which in the quasi-classical approximation is defined as $\tilde{\rho}_0(y) \equiv \rho_0(y(x))$ and can be calculated from the discrete balance equation,

$$\tilde{\beta}_y(y-1)s(y-1)\tilde{\rho}_0(y-1) - \left[\tilde{\alpha}_y^{(th)}(y) + \tilde{\alpha}_y^{(irr)}(y)\right]s(y)\tilde{\rho}_0(y) = 0. \tag{59}$$

In the approximation adopted in Eq. (54), this balance equation, Eq. (59), can be represented in the recurrence form (see details in the Appendix A to Part 1),

$$\frac{\tilde{\rho}_0(y-1)}{\tilde{\rho}_0(y)} \approx S^{-1}\left[\epsilon \exp\left(\frac{2\gamma v}{RkT}\right) + \frac{\xi K_0 R}{D c_y^{(eq)}}\right], \tag{60}$$

where $\epsilon = \exp\left(n_x^{(eq)}\right)$, which in the above considered limits $n_x^{(eq)}, c_y^{(eq)}, \Psi/c_y^{(eq)} \ll 1$ and $R^* \gg 2\gamma v/kT$, ultimately modifies the nucleation rate equation (Eq. (27) from Part 1) to the following form,

$$\dot{N} \approx$$

$$4\pi D R^* \left(S c_y^{(eq)}\right)^2 v^{-1} \left\{\frac{1}{6\pi}\left[\epsilon a x^{*-1} - S^{-1}\frac{\xi K_0}{D c_{e,\infty}}\left(\frac{3v}{4\pi}\right)^{\frac{1}{3}} x^{*-\frac{1}{3}} - S^{-1}\frac{\xi K_0}{D c_{e,\infty}}\left(\frac{3v}{4\pi}\right)^{\frac{1}{3}} a x^{*-\frac{2}{3}}\right]\right\}^{\frac{1}{2}} \exp\left[x^* \ln\frac{S}{\epsilon} - \frac{3}{2}a x^{*\frac{2}{3}} - \frac{3}{4}\frac{\xi K_0}{D c_{e,\infty}}\left(\frac{3v}{4\pi}\right)^{\frac{1}{3}} x^{*\frac{4}{3}}\right]. \tag{61}$$

## 6. Results

At the initial stage of irradiation, which is relatively long (until $\rho_d \ll (4\alpha_{vi}K_0/D_v)^{1/2}$), recombination of point defects is the governing sink for point defects, and thus the vacancy concentration is estimated as $c_v \approx (K_0/\alpha_{vi}D_v)^{1/2}$, where $\alpha_{vi} \approx 6.6 \cdot 10^{20}$ m$^{-2}$ is the point defect recombination constant in bcc Fe (evaluated in [39]), and $D_v$ is the vacancy diffusion coefficient.

Taking into account that in the majority of metals, diffusion is determined by the vacancy mechanism [40], whereas the thermal self-diffusion coefficient in $\alpha$-Fe, $D_s^{(th)}$, and the diffusion coefficient of Cr in Fe-Cr alloys, $D_x^{(th)}$, are rather close [41], i.e. $D_x^{(th)} \sim D_s^{(th)} \approx D_v c_v^{(th)}$, where $c_v^{(th)}$ is the thermal concentration of vacancies, the irradiation-enhanced Cr diffusion coefficient can be estimated as $D \equiv D_x \sim D_v c_v \approx (D_v K_0/\alpha_{vi})^{1/2}$, or $D/K_0 \approx (D_v/\alpha_{vi}K_0)^{1/2}$, and, consequently, the threshold supersaturation $\hat{S}$ will increase with an increase in the damage rate $K_0$, Eq. (57). In particular, this can lead to a situation where, at high damage rates, the concentration $c_y$ is below the threshold value $\hat{c}_y = \hat{S}c_y^{(eq)}$, and thus nucleation of precipitates becomes impossible (as explained above), whereas at lower damage rates $c_y > \hat{c}_y = \hat{S}c_y^{(eq)}$ and thus nucleation and growth of precipitates will occur.

This trend is qualitatively consistent with recent observations of irradiation-induced dissolution of $\alpha'$ precipitates in Fe-15Cr alloys [18], where pre-irradiation samples with precipitates were subjected to self-ion irradiation at different damage rates (from $10^{-5}$ to $10^{-3}$ s$^{-1}$) and temperature $T \approx 673$ K. In these tests, the pre-existing $\alpha'$-precipitates dissolved at high damage rates of $3 \cdot 10^{-4}$ and $10^{-3}$ s$^{-1}$, but at a damage rate of $10^{-4}$ s$^{-1}$ the number density of the precipitates increased and the average radius decreased with increasing dose.

Taking into account that $D_v \approx \alpha f_v a^2 v \exp\left(-\frac{E_v^{(m)}}{kT}\right)$ [40], where $\alpha = 1/6$ is the geometric factor in the bcc lattice, $f_v = 0.727$ is the bcc correlation factor, $a = 0.287$ nm is the lattice parameter for bcc Fe, $v \approx 10^{13}$ s$^{-1}$ is the attempt frequency, and $E_v^m \approx 0.68$ eV is the activation energy of the vacancy migration (estimated, e.g., in [42]), which brings in $D_v(673 \text{ K}) \approx 8 \cdot 10^{-13}$ m$^2 \cdot$s$^{-1}$, the threshold supersaturation can be estimated using Eq. (57) as $\hat{S} \approx 1.2$ at $K_0 = 10^{-5}$ s$^{-1}$, $\hat{S} \approx 1.3$ at $K_0 = 10^{-4}$ s$^{-1}$, and $\hat{S} \approx 1.6$ at $K_0 = 10^{-3}$ s$^{-1}$. After pre-irradiation stage of the tests [18], the Cr concentration in the matrix was $c_y \approx 0.13$, which corresponds to the supersaturation $S = c_y/c_y^{(eq)} \approx 1.44$ (given that, according to the Calphad database, $c_y^{(eq)} \approx 0.09$ at this temperature, cf. [43]). Therefore, nucleation of precipitates will be continued during the irradiation stage at $K_0 = 10^{-5}$ and $10^{-4}$ s$^{-1}$, whereas at $K_0 = 10^{-3}$ s$^{-1}$ it will be suppressed and only dissolution of pre-existing precipitates will take place, in qualitative agreement with observations.

In addition to the formation of fine scale precipitates at $K_0 = 10^{-5}$ and $10^{-4}$ s$^{-1}$, irradiation will also cause pre-existing coarse precipitates to shrink, until complete stabilisation of the system is reached (so-called 'inverse Ostwald ripening'). The stabilised state of the system in this case is characterised by a balance between the dissolution and growth of existing particles in the absence of nucleation of new particles [44], and, as shown in Part 1 ([19]), is described by the thresholds values, $S_{st} = \hat{S}$ and $R_{st} = \hat{R}$, evaluated above in Eqs (57) and (58).

The particle number density in this state is estimated as ([19])

$$N_{st} = \frac{3}{4\pi \hat{R}^3}(S_A - \hat{S}), \tag{62}$$

where $S_A = c_A/c_y^{(eq)}$, and $c_A$ is the total concentration of solute atoms (equal to $\approx 0.16$ in the tests [18]), or, in terms of the dimensionless number density,

$$n_{st} = N_{st}v = \frac{3v}{4\pi \hat{R}^3}(S_A - \hat{S}), \tag{63}$$

as shown in Fig. 1, calculated using Eqs (57), (58) and (63) at $T = 673$ K and $\gamma = 0.2$ J/m$^2$ (presented in Section 2) and other parameters defined above. Correspondingly, stabilisation can be achieved, if $S_A > \hat{S}$, but this may take much longer than the duration of the test.

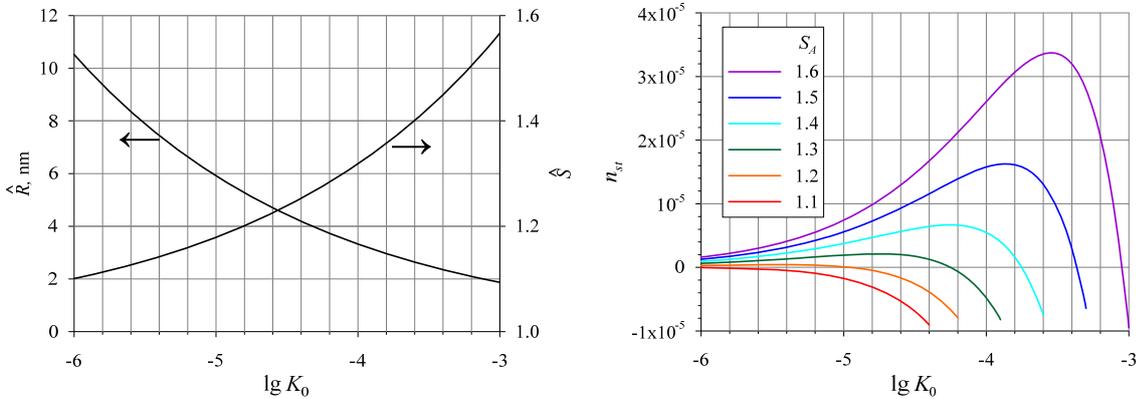

Fig. 1. Threshold radius $\hat{R}$ and supersaturation $\hat{S}$ (*left*), and stabilised particle number density $n_{st}$ (*right*), calculated using Eqs (57), (58) and (63) for $T = 673\ K$ and $\gamma = 0.2$ J/m$^2$.

Given that $\alpha'$ phase precipitates nucleate in the Fe-rich matrix with very little lattice mismatch, allowing them to form a coherent interface, nucleation rate is strongly decreased due to the slow particle growth mechanism (by two-dimensional nucleation of terraces on an atomically smooth particle interface) [45], with its further decrease by ballistic re-solution under irradiation conditions. On the other hand, this decrease can be effectively compensated by a reduction in the nucleation barrier caused by adsorption and mutual annihilation of excess point defects at the coherent particle interface under irradiation conditions [35]. Competition between these effects results in considerable uncertainty regarding the value of the nucleation barrier.

For these reasons, the obtained results should be considered as qualitative estimates that can be improved by replacing $\gamma$ with the effective surface energy and tuning it to better fit the experimental data. This approach is similar to earlier considerations of unirradiated materials, e.g. by Hyland [46] (as applied to nucleation of coherent $Al_3Sc$ precipitates in a dilute Al-Sc alloy), who adjusted the average value of the surface energy $\gamma$ to a significantly higher value compared to existing experimental data and atomistic calculations for this quantity in order to achieve reasonable agreement between the simplified nucleation model (neglecting the above discussed increase in the nucleation barrier for coherent precipitates) and his measurements of nucleation kinetics (cf. [45]).

A kinetic description of irradiation tests could be performed using the model of nucleation and growth kinetics of precipitates that takes into account the ballistic re-solution effects (described by Eqs (61) and (54)), as well as the stabilisation of the system (described by Eqs (57), (58) and (63)), once the model is implemented in a fuel performance code (e.g., BERKUT [47]), as foreseen in the near future.

## 7. Discussion

On the base of a critical analysis of existing models for nucleation of new phase precipitates in binary alloys, such as the generalised Gibbs nucleation model for two-component solid solutions considered in the regular solution approximation [8, 9], a new kinetic model of homogeneous nucleation and growth of $\alpha'$-phase precipitates in Fe-Cr alloys was developed. The simplifications of the regular solution model was overcome by a more realistic approach, in which each of the two phases is characterised by its own (independent of the second phase) set of microscopic parameters, associated with the actual position of the equilibrium lines (binodals) on the binary phase diagram. For a diluted Fe-Cr system with low mutual solubility of the two components in both phases (i.e. Cr in the matrix $\alpha$ phase and Fe in the precipitated $\alpha'$ phase), the weak solution approximation was applied to each phase, facilitating the analytical calculation of the composition and size of the critical nucleus.

In this approach, the basic postulate of the binary nucleation theory formulated by Hobstetter [10] and Reiss [11] was used to calculate the critical nucleus size and composition (and the related nucleation barrier), which for binary systems corresponds to a saddle point on the free energy surface of cluster formation. It was shown that the critical nucleus composition belongs to the stable zone of the $\alpha'$ phase on the Fe-Cr binary phase diagram, and penetrates even deeper into the stable zone if the dependence of the interface energy $\gamma$ on the precipitate composition is taken into account (contrary to the predictions of the earlier model [23]), while the contribution of elastic strain energy to this effect is insignificant.

It was also noted that the kinetics of nucleation of binary precipitates cannot be reduced to simple predictions for unary precipitates, but should be analysed within the framework of the Reiss kinetic theory of binary nucleation [11], modified by Langer [15], who showed that the evolution of the nucleus through a two-dimensional free energy barrier occurs in the direction of an unstable mode at the saddle point. This approach was implemented in this work with regard to the Fr-Cr alloy, and further simplified using the so called 'quasi-classical' approximation, which correctly takes into account the direction of nucleus evolution through the nucleation barrier, while conservatively (under)estimating the pre-exponential kinetic factor of the nucleation rate.

To determine the direction of nucleus evolution, the change in the composition of the supercritical nucleus during its growth was analysed and compared with existing experimental observations from the literature. In particular, it was shown that the composition of the growing particle still belongs to the stable zone of the phase diagram and approaches the equilibrium (binodal) concentration of the $\alpha'$ phase. This model prediction is qualitatively consistent with the results of Mössbauer spectroscopy in long term ageing tests [26] and the results of APT in irradiation tests [28], but contradicts observations in short ageing tests on Fe-20%Cr alloys [29].

A possible reason for the discrepancy between the test results were discussed in [28], where it was associated with a very strong magnification effect of APT for small nanoscale precipitates formed during short ageing times. Another possible reason for the discrepancy of the model predictions with observations in the tests [29] may be related to the high degree of oversaturation of these alloys, when the Gibbs free energy of nucleus formation cannot be described by the weak solution approximation used in the model. Moreover, classical nucleation theory may not be sufficient for such systems with a high degree of metastability. It may be necessary to apply van der Waals theory, which takes into account the dependence of thermodynamic parameters on concentration gradients (as discussed in [8] and considered in phase field calculations [33], which were successfully applied to analyse the tests [29]).

Following the general approach developed in Part 1 ([19]), the model was further modified to account for the influence of ballistic re-solution on the kinetics of nucleation and growth of $\alpha'$ precipitates. It was shown that nucleation of precipitates is not possible if the supersaturation ratio $S$ of the alloy is below the threshold value, $1 \leq S < \hat{S}$, whereas at higher supersaturations, irradiation causes existing coarse precipitates to shrink, but at the same time, fine scale precipitates form and grow until the system is completely stabilised (so-called 'inverse Ostwald ripening'). The stabilised state of the system is established when a balance is reached between the dissolution and growth of existing particles in the absence of nucleation of new particles, and is characterised by the threshold values, $S_{st} = \hat{S}$ and $R_{st} = \hat{R}$, evaluated by the model. Since the supersaturation threshold depends on the irradiation damage rate $K_0$ (at the initial, relatively long stage of irradiation), different regimes (with $S > \hat{S}(K_0)$ and $S < \hat{S}(K_0)$) can be realised in the same alloy at different damage rates, in qualitative agreement with recent observations [18]. A quantitative kinetic description of these tests could be performed using the new model, once it is implemented in the fuel performance code (as foreseen in the near future).

## 8. Conclusions

A new kinetic model of homogeneous nucleation of $\alpha'$-phase precipitates in diluted Fe-Cr alloys was developed using the weak solution approximation (applicable to both the matrix and precipitate phases), in which simplifications of the regular solution models [8, 9] were avoided. The nucleation kinetics was analysed within the framework of the Reiss kinetic theory of binary nucleation and further simplified using the 'quasi-classical' approximation, which correctly takes into account the direction of nucleus composition evolution through the nucleation barrier, while conservatively (under)estimating the kinetic factor of the nucleation rate.

It was shown that the composition of the critical nucleus belongs to the stable zone of the $\alpha'$ phase on the Fe-Cr binary phase diagram, and penetrates even deeper into the stable zone if the dependence of the interface energy $\gamma$ on the precipitate composition is taken into account, while the contribution of elastic strain energy to this effect is insignificant.

The new model predicts that, as the nucleated particle grows, its composition approaches the equilibrium (binodal) concentration of the $\alpha'$ phase, which is consistent with the results of long term ageing and irradiation tests, but contradicts APT measurements in short term ageing tests on Fe-20%Cr alloys. This discrepancy can be explained by a very strong magnification effect in APT for small nanoscale precipitates (as suggested in [28]), or, alternatively, by the relatively high Cr concentration of the alloy in these tests, where the system cannot be described using the weak solution model.

The model was further modified to account for the influence of ballistic re-solution on the kinetics of nucleation and growth of $\alpha'$ precipitates, following the general approach proposed in Part 1 ([19]). It was shown that nucleation of precipitates is unfeasible if the supersaturation ratio $S$ of the alloy is below the threshold value $\hat{S}$, which depends on the irradiation damage rate $K_0$, and therefore, different regimes of precipitation behaviour (nucleation and growth at $S > \hat{S}(K_0)$, or dissolution at $S < \hat{S}(K_0)$) can occur in the same alloy at different damage rates, in qualitative agreement with recent observations [18].

**Acknowledgements**

Dr. V. Tarasov (IBRAE, Moscow) is acknowledged for valuable discussions and assistance in numerical calculations (Fig. 1).

**Appendix**

The elements $D_{ij} = \frac{1}{2} \frac{\partial^2 \Delta G_0(x,y)}{\partial x_i \partial x_j}\Big|_{x^*,y^*}$ of the matrix **D** calculated to the principal term in the expansion by $n_x^{(eq)} \ll 1$ using Eq. (3) take the form,

$$D_{11} \equiv D_{xx} \approx \frac{kT}{2}\left(\frac{1}{x^*} - \frac{2}{9}\frac{\gamma}{kT}v^{\frac{2}{3}}y^{*-\frac{4}{3}}\right) \approx \frac{(kT)^4}{64\pi\gamma^3 v^2}\left(\ln S - n_x^{(eq)} + \frac{n_x^{(eq)}}{S}\right)^3 \left(\frac{3S}{n_x^{(eq)}} - 3 - \ln S + n_x^{(eq)}\right), \text{(A.1)}$$

$$D_{22} \equiv D_{yy} \approx \frac{kT}{2}\left(\frac{x^*}{y^{*2}} - \frac{2}{9}\frac{\gamma}{kT}v^{\frac{2}{3}}y^{*-\frac{4}{3}}\right) \approx \frac{(kT)^4}{64\pi\gamma^3 v^2}\left(\ln S - n_x^{(eq)} + \frac{n_x^{(eq)}}{S}\right)^3 \left(2\frac{n_x^{(eq)}}{S} - \ln S + n_x^{(eq)}\right), \text{(A.2)}$$

$$D_{12} \approx \frac{kT}{2}\left(\frac{1}{y^*} - \frac{2}{9}\frac{\gamma}{kT}v^{\frac{2}{3}}y^{*-\frac{4}{3}}\right) \approx -\frac{(kT)^4}{64\pi\gamma^3 v^2}\left(\ln S - n_x^{(eq)} + \frac{n_x^{(eq)}}{S}\right)^3 \left(3 + \ln S - n_x^{(eq)}\right), \quad \text{(A.3)}$$

where $D_{11} > 0$ and $D_{12} < 0$ above the critical point, $\ln S - n_x^{(eq)} + \frac{n_x^{(eq)}}{S} > 0$ (see Eq. (14)), while $D_{22}$ changes its sign, and is positive for relatively small oversaturations, $S \leq 1.25$ (at $n_x^{(eq)} \approx 0.1$),

$$\det \mathbf{D} = D_{11}D_{22} - D_{12}^2 \approx -\frac{2}{9}kT\sigma v^{\frac{2}{3}}y^{*-\frac{4}{3}}\left(\frac{1}{x^*} + \frac{2}{y^*} + \frac{x^*}{y^{*2}}\right) \approx -\frac{2}{9}kT\sigma v^{\frac{2}{3}}y^{*-\frac{4}{3}}\frac{1}{x^*} \approx$$

$$-3\left(\frac{1}{32\pi\gamma^3 v^2}\right)^2 (kT)^8 \left(\ln S - n_x^{(eq)} + \frac{n_x^{(eq)}}{S}\right)^7 \frac{S}{n_x^{(eq)}} < 0, \quad \text{(A.4)}$$

$$d_a = -D_{11}/D_{12} = \frac{3S}{n_x^{(eq)}\left(3 + \ln S - n_x^{(eq)}\right)}, \quad \text{(A.5)}$$

$$d_b = -D_{22}/D_{12} = \frac{\left(2\frac{n_x^{(eq)}}{S} - \ln S + n_x^{(eq)}\right)}{\left(3 + \ln S - n_x^{(eq)}\right)}, \quad \text{(A.6)}$$

$$r = \frac{\beta_x^*}{\beta_y^*} \approx \frac{(1-c_y)}{c_y}\frac{R^*}{a} \gg 1, \quad \text{(A.8)}$$

where $\beta_y^* = \beta_y(R^*) = Dc_y R^{*-1} v^{-1}$, and $\beta_x^* = \beta_x(R^*) \approx D(1-c_y)a^{-1}v^{-1}$ (see Eqs. (26) and (27)).

Substituting these expressions into Eqs (35)–(40), an analytical formula for the nucleation rate $\dot{N}$ is obtained, which, however, turns out to be rather cumbersome and therefore can be further analysed numerically.